\documentclass[pre,twocolumn,superscriptaddress]{revtex4-1}
\usepackage{graphicx,epsfig}
\usepackage{epstopdf}
\usepackage{natbib}
\usepackage{rotating}
\usepackage{subfigure}
\usepackage{booktabs}
\DeclareGraphicsExtensions{. jpg,. pdf, . mps, .png, .eps, . ps, . EPS}

\usepackage{amsmath}
\usepackage{color}
\begin{document}
\newcommand{\beq}{\begin{equation}}
\newcommand{\eeq}{\end{equation}}
\newcommand{\bea}{\begin{eqnarray}}
\newcommand{\eea}{\end{eqnarray}}
\newcommand{\gt}{\tilde{g}}
\newcommand{\mt}{\tilde{\mu}}
\newcommand{\et}{\tilde{\varepsilon}}
\newcommand{\ct}{\tilde{C}}
\newcommand{\bt}{\tilde{\beta}}
\newcommand{\angstrom}{\mbox{\normalfont\AA}}

\newcommand{\avg}[1]{\langle{#1}\rangle}
\newcommand{\Avg}[1]{\left\langle{#1}\right\rangle}
\newcommand{\cor}[1]{\textcolor{red}{#1}}

\title{Network measures for protein folding state discrimination}
\author{Giulia Menichetti}
\affiliation{Department of Physics and Astronomy and INFN Sez. Bologna, Bologna University, Viale B. Pichat 6/2 40127 Bologna, Italy}
\author{Piero Fariselli}
\affiliation{Department of Informatics Science and Engineering, Bologna University
Mura Anteo Zamboni 7 40127 Bologna, Italy}
\author{Daniel Remondini}
\affiliation{Department of Physics and Astronomy and INFN Sez. Bologna, Bologna University, Viale B. Pichat 6/2 40127 Bologna, Italy}

\begin{abstract}
Proteins fold using a two-state or multi-state kinetic mechanisms, but up to now there isn't a first-principle model to explain this different behaviour. 
We exploit the network properties of protein structures by introducing novel observables to address the problem of classifying the different types of folding kinetics.
These observables display a plain physical meaning, in terms of vibrational modes,  possible configurations compatible with the native protein structure, and folding cooperativity.
The relevance of these observables is supported by a classification performance up to $90\%$, even with simple classifiers such as disciminant analysis.

\end{abstract}

\maketitle

\section{Introduction}

Protein folding is one of the most studied biophysical problems \cite{Tramontano2005}, and despite the fact that protein folding is a straightforward biophysical process \citep{Anfinsen1961}, up to now there is not a general agreement on how and why proteins fold \citep{EnglanderMayne2014}. 
Experimentally, protein folding kinetics is divided into two fundamental categories: Two-State (TS) folding and Multi-State (MS) folding. 
While Two-State kinetics can be considered as an "all-or-none" transition, Multi-State folding displays at least one or more intermediates.
Measuring experimentally the type of protein kinetics is not an easy task \citep{EnglanderMayne2014}, and computational studies can help unraveling relevant mechanisms \citep{ChangBriefBio2014}.
The classification of proteins in these two major groups and the related prediction of folding rates have been widely debated in recent years. 
Previous studies have focused on several different types of predictors \citep{Huang2008, Huang2012, Song2010a}, exploiting the main features of protein primary structures and protein contact map representations (for a review see  \citep{EnglanderMayne2014}). 
The geometry of the native protein structure plays a relevant role to infer the value of the folding rate. For this task, different predictors have been proposed based on: structural topology measures such as contact order \citep{Plaxco2000, kfold} and long range contact order \citep{Gromiha2001}, clustering coefficient, characteristic path length and assortativity coefficient \citep{Song2010a}, cliquishness \citep{Micheletti2003}, chain length and amino acid  composition \citep{Huang2008, Huang2012}. 
These observables or combinations of them were usually evaluated by means of binary logistic regression (BLR) and support vector machine (SVM). 
In particular, SVM classifiers map the data into a higher dimensional feature space, that is usually not easily interpretable in terms of the original variables.
Most predictors do not usually perform in the same way both for Two-State and Multi-State proteins, causing unbalanced value of sensitivity and sensibility according to the target of the analysis.

In this paper we predict if a protein behaves as Two-State or Multi-State using only the native structure. As made before by other authors \citep{Vendruscolo2002, DiPaola2013a, Bartoli2007}, we represent the protein 3D structure as a contact map between amino-acid residues (Protein Contact Network PCN). The PCN is the adjacency matrix of a graph, whose links represent the contacts between residues. Our assumption is that the native PCN contains a clue of the protein folding kinetics. In this respect, we introduce three observables that should take into account that Multi-State proteins must be trapped into one or more intermediate states. 
First of all, we make the hypothesis that MS protein should have more configurational microstates to explore than the TS proteins, and we implemented a measure of Network Entropy to quantify this aspect. 
Second, from a modified version of the PCN, that keeps only long-range contacts but preserving network connectivity, we evaluated the spectrum of the Laplacian matrix, since it has been shown that its vibrational properties can be used to model experimental data \citep{Bahar1997,Maritan2004}.
Finally, in order to measure the folding cooperativity \citep{Dill1993}, we evaluate the fraction of sequence separation (diagonals of the PCN) that do not contain residue contact pairs. 
The rationale of this measure is that the more diffuse is the cooperation (most of the diagonal participate) the less probable is to be trapped in intermediate states.
In order to keep these observables as independent as possible from the protein size, they were accordingly rescaled by a function of residue chain length.
In this paper we show that these observables perform very well even with a simple discriminant classifier, that allows to give a intuitive biophysical interpretation to our results.


\section{Definition of the observables}

\begin{figure*}
\center
\includegraphics[width=\textwidth]{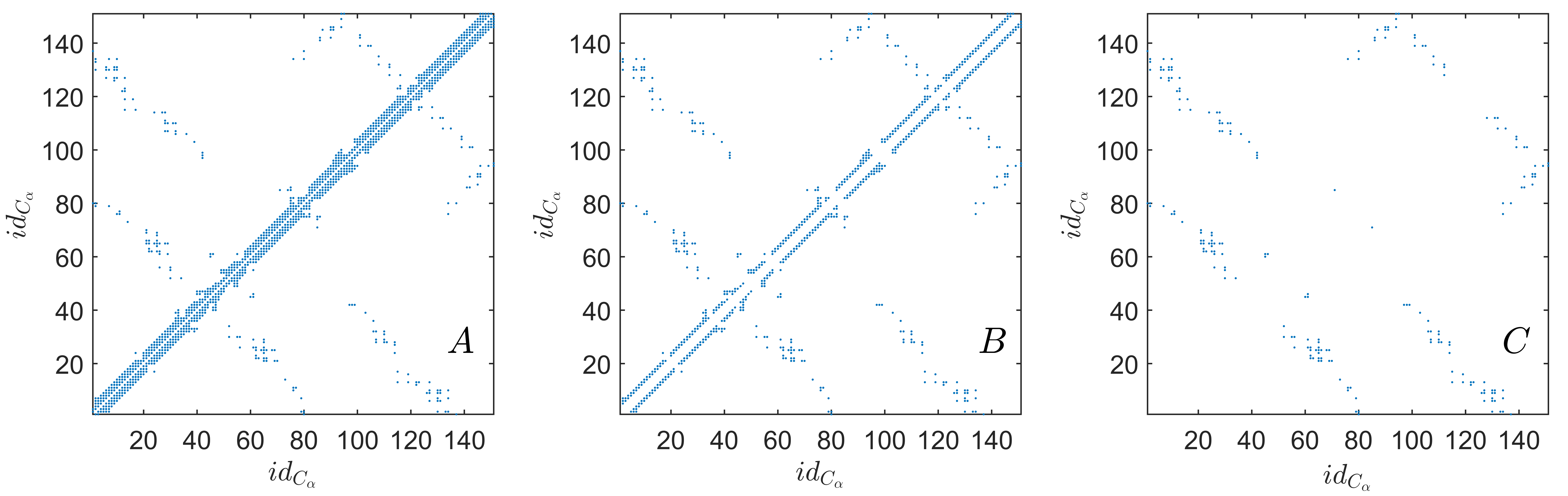}
\caption{Different representations for protein $1A6N$, with Multi-State folding kinetic and  151 $C_{\alpha}$. In Panel A the whole PCN is displayed. In Panel B, once calculated the parameter $b=4$ ( $b$ defines the number of diagonals needed to break the protein network in more than one component), 3 diagonals were removed in the upper triangular and in the lower triangular. In Panel C the related Long-range Interaction Network (LIN) is shown\citep{Gromiha2001, Song2010a}.}
\label{contactmappanel3}
\end{figure*}

\subsection{Entropy-based measure $S_R$}

The first observable introduced is associated with the-so called Entropy of a network ensemble \citep{Park2004}.
Network entropy is related to the logarithm of the number of typical networks (in our case the possible PCNs) that satisfy some given constraints based on node and link features of a real network instance  (the studied protein).
We hypothesize that the network structure of the protein native state retains information related to the protein folding process (such as the possible intermediate states that could be represented as non-native PCNs).
It has been recently applied in a biological context, as a measure of the “parameter space” available to the cell (in terms of gene expression profile or clonal diversity) and it allowed to successfully characterize different cell states related to different cancer stages or to physiological ageing \citep{Menichetti2015}. 
In our approach, each protein is considered as an undirected weighted network, in which we integrate the information on the topological structure given by protein contacts with the information on residue interactions given by $\{P_{ij}\}$ weights (related to the contact potential matrix $M$ described in \cite{Bastolla2001} and explained in details in the ''Experimental Data'' Section, Eq. \ref{ContactPotential}). 
 
For each protein, we calculated the Network Entropy $S_{BS}$ for two different ensembles, with a different number of constraints: in the first ensemble, we only fix the strength sequence $\{s_i\}$ of the protein network ($S_s$), while in the second ensemble ($S_{ks}$) we fix both strength sequence $\{s_i\}$ and degree sequence  $\{k_i\}$. 
The degree sequence $\{k_i \}$ and the strength sequence $\{s_i \}$ are respectively defined as the number of contacts and the weighted sum of contacts of the nodes in the network. 

Network Entropy can be generally defined as
\begin{equation}
\label{entropy}
S=-\sum_{i<j}\sum_{w=0}^{\infty}\pi_{ij}(w)\log (\pi_{ij}(w)).
\end{equation}
where, for the sake of simplicity, weights $w$ are discrete and $\pi_{ij}(w)$ is the probability to observe weight $w$ between residue $i$ and $j$.
The  constraints previously defined for the calculation of maximum network entropy  are written as
\begin{align}
s_i^{prot}&= \sum_j \sum_w w \pi_{ij}(w)\quad \forall i \\
k_i^{prot}&= \sum_j \sum_{w \ne 0} \pi_{ij}(w)\quad \forall i,
\end{align}
where the average values of strength sequence and degree sequence over the network ensemble are enforced to match the real features of the selected protein, i.e. $\{s_i^{prot}\}$ and $\{k_i^{prot}\}$.
The network entropy observable $S_R$ is defined as the ratio between the two entropies with a different number of constraints
\begin{equation}
S_R = S_s/S_{ks}
\label{S_R}
\end{equation}
with $S_s\ge S_{ks}$ given the fewer number of constraints.
The closer the value of $S_R$ to $1$, the less relevant is the role of the degree sequence constraint $\{k_i\}$, and thus most of the information on possible PCN configurations is enclosed in the strength sequence $\{s_i\}$ only. 
On the contrary, a large value of $S_R$ implies that the given strength sequence is compatible with a larger number of degree sequences (corresponding to more possible PCNs).
Thus, MS proteins could in principle have larger $S_R$ values than TS proteins: having more stable (or metastable) configurations available could be reflected in a larger number of available configurations as measured by $S_R$.


\subsection{Laplacian-based observables $\lambda_i$}

The Laplacian operator $L$ on networks is a positive semi-definite operator that plays a major role in the study of diffusion processes on networks, in node clustering and network visualization \citep{Chung1994,Girvan2002}, and it has already been applied to characterize protein features \citep{Burioni2004}. 
Given an adjacency matrix $A$ without self loops, we define the Laplacian operator as
\begin{equation}
L = K-A; \quad K_{ij} = k_i\cdot\delta_{ij}
\end{equation}
Remarkably, in case of a N-lattice network, the eigenvalue problem for the Laplacian operator can be put in analogy with the discretization of an N-dimensional elastic membrane \citep{Biyikoglu2007}.
With this analogy in mind, the eigenvalues of the Laplacian matrix can be associated with the oscillating frequencies (harmonics) of the vibrating modes on the membrane, with the largest eigenvalues corresponding to the highest frequencies.

Since we suppose that the relevant information on folding kinetics can be contained in the long-range contacts of the native folded state \citep{Robson1993}, we decided to partially remove the backbone contacts from the original PCN.
In more detail, for each protein we evaluated $b$, the number of $d-$diagonals (the set of links between nodes at a distance of $d$ residues along the backbone, see Eq. \ref{ddiag}) needed to break the protein network in more than one component. 
Then, $b-1$ diagonals were removed from the PCN. 
We remark that this procedure is specific for each protein, i.e. the parameter $b$ depends on the PCN of each protein, and moreover, once removed $b-1$ diagonals, the PCN is still connected, thus generating a unique eigenvalue spectrum for $L$. 
Also other authors considered a reduced PCN \cite{Gromiha2001, Song2010a}, but they considered a unique threshold to define long-range interactions, considering only inter-residue distances $d > 12$ independently from protein size and structure (see Fig. \ref{contactmappanel3}).
Once the laplacian spectrum of this modified version of PCN was computed for each protein, we considered as observables $\{\lambda_i\}$, the largest eigenvalues of $L$ rescaled by the number of residues $N_C$. 
This rescaling was chosen because the highest eigenvalue has $N_C$ as upper bound, so our observables could in principle be linearly dependent on the number of residues.
According to the vibrational interpretation of the Laplacian, these eigenvalues represent the highest vibrational frequencies associated to the long-range structure of the protein.

\subsection{Inter-residue link density $R_0$}
Each PCN is an adjacency matrix, in which the d-diagonals
\bea
\label{ddiag}
PCN_{ij}, \forall i,j: |i-j|=d
\eea
contain all the links between residues with a sequence separation equal to $d$ (ranging from $1$ to $N_C-1$) with respect to the protein backbone. 
The observable $R_0$ is defined as the ratio between the number of d-diagonals without links and the number of residues $N_C$ of the protein, i.e.
\begin{equation}
R_0= \frac{1}{N_C}\sum_d \delta \left (\sum_{i<j, |i-j|=d}PCN_{ij},0 \right)
\end{equation}
where $\delta$ is the Kronecker delta.
For a given protein, a low value of $R_0$ implies that the residues interact at many different levels of sequence separation (different values of $d$). 
On the contrary, a high value of $R_0$ indicates that, in such protein structure, the residue interactions are more localized and show less cooperativity.

\section{Experimental data}

The proteins to be classified as TS or MS are obtained from the manually-annotated dataset curated by Ivankov and Finkelstein \citep{Ivankov2004a}. 
The dataset consists of 63 proteins, 25 of which are classified as Multi-State and 38 as Two-State.
The protein structures are taken from the Protein Data Bank (\textit{www.rcsb.org}). 
We model the protein structure with its alpha carbon ($C_{\alpha}$) trace. We collapse the entire protein structures into related contact matrices between the $C_{\alpha}$'s of the residues. 
Contact matrices represent a common way of modeling proteins, that guarantees a good representation of the complex relationship between structure and function of proteins, while cutting out the redundant information embedded in the whole 3D structure. 
Contact matrices  are essentially networks in which the role of nodes is played by residues and edges or ``contacts" depend upon a notion of ``distance" between each couple of residues. 
The position of an entire amino acid is usually collapsed into the corresponding $C_{\alpha}$ and the ordering of nodes is physically justified by the primary structure of the protein, i.e. the protein backbone. The backbone is composed by residues that are in sequence and whose distance ranges 3-4 $\angstrom$, the so-called ``peptide bond". 
Once obtained the $C_{\alpha}$ spatial distribution, the contact matrix $D$ is considered, where each element $d_{ij}$ is the 3D Euclidean distance between the $i^{th}$ and $j^{th}$ residues. 
The protein contact network PCN is then obtained by choosing an upper threshold of 8 $\angstrom$ \citep{Aftabuddin2006, Barah2008, Bagler2005,Brinda2005}: 
\begin{equation}
PCN_{ij} = 1 \quad if \quad d_{ij}<8\angstrom
\end{equation}

In order to build our network-based observables, we retrieved data regarding amino acidic interactions, such as hydropathy indexes \citep{kytedoolittle, miyazawajernigen} and contact potentials, namely, $20\times 20$ matrices describing the interactions between the 20 side-chains \citep{Bastolla2001, BETM990101, LIWA970101}. 
Each element of the contact potential represents the interaction strength between a pair of amino acids at contact. 
In this paper we provide results only for the contact potential matrix $M$ described in \citep{Bastolla2001}, since the results obtained with other potentials \cite{BETM990101, LIWA970101} are very similar.  
For each protein, starting from the known residue sequence, we obtain the contact potential matrix $P$, in which
\bea
P_{ij}&=& M(r_i, r_j)
\label{ContactPotential}
\eea
where $r_i$ is the amino acid residue corresponding to the $i^{th}$ $C_{\alpha}$, and the matrix $P$ has the same size of the related PCN.
Since $P_{ij}$ can assume negative values, in order to have only positive weights we shifted their values in each PCN to have the smallest weight equal to one.

\section{Results}

\subsection{Protein classification by Discriminant Analysis}

\begin{figure*}
\center
\includegraphics[width=\textwidth]{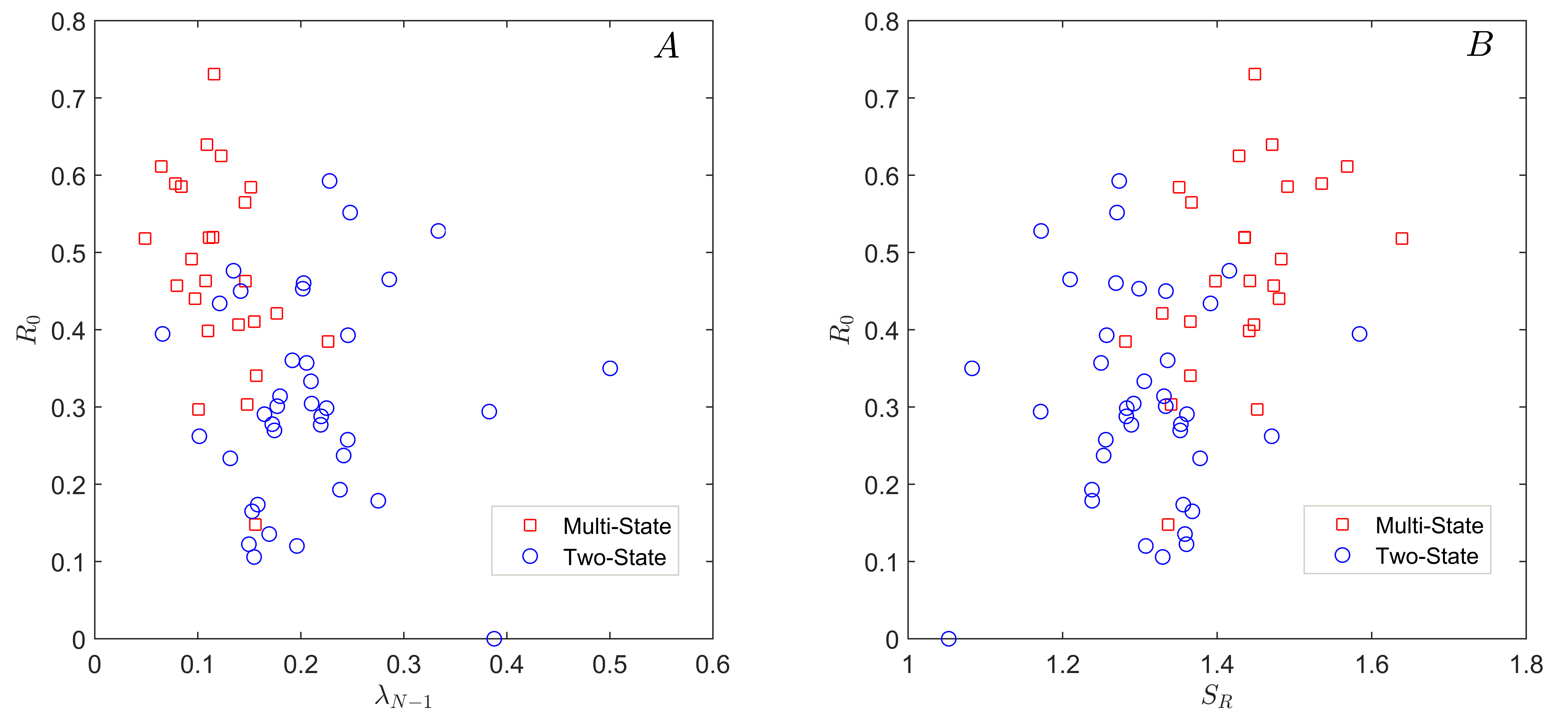}
\caption{Scatterplots of the top-ranking classification couples. 
Panel A (left): $\lambda_{N-1}$ and $R_0$ (classification $=88.33\%\pm 1.10\%$). 
Panel B (right): $S_R$ and $R_0$ (classification $=84.50\%\pm 1.30\%$). }
\label{panel_class}
\end{figure*}

All the defined observables were considered as features for classification of TS and MS proteins. 
Fisher Discriminant Analysis, one of the most robust classifiers that allow simple interpretations of the obtained classes, was applied to single observables and to their combinations (i.e. couples, triplets and quadruplets of observables).
A 10-fold crossvalidation with 10000 resamplings was used to assess the performance of our classifiers, that we will describe as the average value over the resamplings and with the standard deviation as the confidence interval. 
Given the presence of homologous proteins, in each partition of the 10-fold crossvalidation all the homologous proteins were kept together to reduce the risk of overestimating the classifier perfomance.
In order to characterize the homogeneity of the classification performance over both TS and MS classes, we consider the Matthews Correlation Coefficient MCC, defined as 
\begin{equation}
MCC=\frac{T_P\cdot T_N - F_P \cdot F_N}{\sqrt{(T_P+F_P)(T_P+F_N)(T_N+F_P)(T_N+F_N)}}
\end{equation}
where $T_P$ is the number of true positives, $T_N$ the number of true negatives, $F_P$ the number of false positives and $F_N$ the number of false negatives.
A coefficient of $+1$ defines a perfect prediction, $0$ is nothing more than a random prediction, while $-1$ reflects total disagreement between prediction and observation.\\
Classification with the novel observables was drastically improved, with performances up to $88\%$ and $MCC= 0.76$, indicating that both classes are correctly classified at high level (previous results in \cite{kfold} with SVM classifier were around $80\%$).
The details of all performances, both for single observables and their combinations in couples, are shown in Tab. \ref{NEWO}.
The combinations of the largest Laplacian eigenvalues $\lambda_{N}, \lambda_{N-1}, \lambda_{N-2}$ with the link density $R_0$ produce the best performances of classification, with a top-score value given by the couple  $(\lambda_{N-1},R_0)$, with $88.33 \%\pm 1.10\%$ correctly classified proteins, with a highly homogeneous performance on both classes ($87.20 \% \pm 2.21\%$ MS  and $89.07 \% \pm 1.01\%$ TS classified proteins, $MCC=0.76 \pm 0.02$). 
The entropy ratio $S_R$ is the best single classifier ($80.36 \%\pm 1.81\%$ correctly classified proteins, $MCC=0.59 \pm 0.04$ ), and it also has a very high performance in combination with $R_0$ ($84.50 \%\pm 1.30\%$ correctly classified proteins, $MCC=0.67 \pm 0.03$). 
Classifying without cross-validation, i.e. using the entire set of  $(\lambda_{N-1},R_0)$ features,  we obtain a  performance of $90.48 \%$, with $92.00\%$ for MS proteins and $89.47\%$ for TS proteins.
We also considered  higher-dimensional signatures (with combinations of 3 and 4 observables) but the performance was not significantly increased.
In Fig. \ref{panel_class} we show the scatter-plot for two top-scoring couples: $(\lambda_{N-1},R_0)$ and $(S_R,R_0)$.
As it can be seen, the two classes are almost linearly separated, and this may allow a simple interpretation in terms of the observables, see Discussion.

Since in previous studies \citep{Huang2008} it has been shown that the chain length $N_C$ is a good classifier of folding classes, we rescaled our observables in order to keep them as much independent as possible from protein length.
Moreover, as a comparison for classification performance, we used $N_C$ as a variable for discrimination.
In our dataset, the $N_C$ parameter correctly classifies $78.23\% \pm 1.52$ proteins, with a large unbalance between correctly classified proteins from the two classes: $57.61\% \pm 3.04$ for MS proteins and $91.80\% \pm 1.29$ for TS proteins ($MCC= 0.54 \pm 0.03$).
Hydrophobic force has  always been indicated as one of the major drivers for the protein folding \cite{DillMacCallumm2012}. 
Since each amino acid is associated with a hydropathy index $h_i$, a number representing the hydrophobic or hydrophilic properties of its side-chain,  each protein can be associated to an average hydrophobicity value $\avg{h}$:
\begin{equation}
\avg{h}=\frac{1}{N_C}\sum_i h_i^{KD}
\end{equation}
where $h_i^{KD}$ refers to the hydropathy index of residue $i$ when the Kyte-Doolittle (KD) scale \citep{kytedoolittle} is considered. 
The average hydrophobicity $\avg{h}$ has been often coupled to $N_C$ to classify the protein folding kinetics. 
In the considered dataset the couple $(N_C,\avg{h})$ correctly classifies $73.71\% \pm 2.15$ proteins, with $MCC=0.44 \pm 0.05$.\\
We also considered the classification power of structural topology measures such as contact order \citep{Plaxco2000, kfold}
\begin{equation}
CO=\frac{1}{N_C L_C}\sum_{ij}^{N_C}PCN_{ij}\cdot|i-j|
\end{equation}
where $L_C$ is the total number of contacts for the given PCN, and long range contact order \citep{Gromiha2001}
\begin{equation}
LRCO=\frac{1}{N_C^2}\sum_{ij, |i-j|>12}^{N_C}PCN_{ij}\cdot|i-j|
\end{equation}
Both these measures perform poorly: CO correctly classifies $68.42 \% \pm 0.65$ proteins with $MCC=0.36 \pm 0.01$ while LRCO guesses right $54.38 \% \pm 2.41$ proteins with $MCC=0.14 \pm 0.05$. 
A complete summary of the performances of classical measures (both singularly and in couples) can be found in Tab. \ref{OLDO}.\\

\begin{table*}[t]
{\centering
\subtable[ Classification performances]{%
\begin{tabular}{l|l|l|l|l|l}
\multicolumn{1}{c|}{$\%$} &
\multicolumn{1}{c}{$\lambda_N$} & \multicolumn{1}{|c}{$\lambda_{N-1}$} & \multicolumn{1}{|c}{$\lambda_{N-2}$} & \multicolumn{1}{|c}{$R_0$} & \multicolumn{1}{|c}{$S_R$}\\
\hline
$\lambda_N$ &76.6 $\pm$ 1.3	&74.7 $\pm$ 1.4	&70.7 $\pm$ 1.8	&85.2 $\pm$ 1.4	&78.4 $\pm$ 1.2\\
\hline
$\lambda_{N-1}$ & \quad	&76.7 $\pm$ 1.4	&70.7 $\pm$ 1.4	&\textbf{88.3} $\pm$ 1.1	&75.9 $\pm$ 2.3\\
\hline
$\lambda_{N-2}$ &\quad	&\quad 	&77.6 $\pm$ 1.1	&87.3 $\pm$ 1.4 &76.5 $\pm$ 1.9\\
\hline
$R_0$ &\quad & \quad & \quad &75.9 $\pm$ 1.2	&84.5 $\pm$ 1.3 \\
\hline
$S_R$ & \quad & \quad & \quad & \quad &80.4 $\pm$ 1.8\\
\end{tabular}
}} \quad
{\centering
\subtable[ Matthews correlation coefficient $MCC$]{%
\begin{tabular}{l|l|l|l|l|l}
\multicolumn{1}{c|}{$\quad$} &
\multicolumn{1}{c}{$\lambda_N$} & \multicolumn{1}{|c}{$\lambda_{N-1}$} & \multicolumn{1}{|c}{$\lambda_{N-2}$} & \multicolumn{1}{|c}{$R_0$} & \multicolumn{1}{|c}{$S_R$}\\
\hline
$\lambda_N$ &0.57 $\pm$ 0.02	&0.53 $\pm$ 0.02	&0.46 $\pm$ 0.03	& 0.69 $\pm$ 0.03	&0.60 $\pm$ 0.02\\
\hline
$\lambda_{N-1}$ & \quad	&0.58 $\pm$ 0.02	& 0.46 $\pm$ 0.03	& \textbf{0.76} $\pm$ 0.02	& 0.57 $\pm$ 0.04\\
\hline
$\lambda_{N-2}$ & \quad	& \quad & 0.59 $\pm$ 0.02	& 0.74 $\pm$ 0.03 & 0.56 $\pm$ 0.04\\
\hline
$R_0$ & \quad	& \quad 	& \quad &0.52 $\pm$ 0.02	& 0.67 $\pm$ 0.03 \\
\hline
$S_R$ & \quad	& \quad	& \quad & \quad 	& 0.59 $\pm$ 0.04\\
\end{tabular}
}
}
\caption{Classification performances of the newly defined observables and Matthews correlation coefficient with quadratic discriminant analysis.
The tables show the performances of couples of observables, with the performance of the single observables along the diagonal; the best performance is bold-typed.
The results presented are the average values of 10-fold cross-validation over 10000 instances and their standard deviation.}\label{NEWO}
\end{table*}

\begin{table*}[t]
{\centering
\subtable[ Classification performances]{%
\begin{tabular}{l|l|l|l|l}
\multicolumn{1}{c|}{$\%$} &
\multicolumn{1}{c}{$N_C$} & \multicolumn{1}{|c}{$\Avg{h}$} & \multicolumn{1}{|c}{$CO$} &  \multicolumn{1}{|c}{$LRCO$}\\
\hline
$N_C$ &78.2 $\pm$ 1.5	& 73.7 $\pm$ 2.1	&78.4 $\pm$ 1.7	& \textbf{80.0} $\pm$ 1.1\\
\hline
$\Avg{h}$ & \quad	& 57.3 $\pm$ 1.4	&72.6 $\pm$ 2.8	& 62.7 $\pm$ 1.9\\
\hline
$CO$ &\quad	&\quad 	&68.4 $\pm$ 0.7	&70.4 $\pm$ 2.1\\
\hline
$LRCO$ &\quad & \quad & \quad &54.4 $\pm$ 2.4\\
\end{tabular}
}} \quad
{\centering
\subtable[ Matthews correlation coefficient $MCC$]{%
\begin{tabular}{l|l|l|l|l}
\multicolumn{1}{c|}{$\quad$} &
\multicolumn{1}{c}{$N_C$} & \multicolumn{1}{|c}{$\Avg{h}$} & \multicolumn{1}{|c}{$CO$} &  \multicolumn{1}{|c}{$LRCO$}\\
\hline
$N_C$ &0.54 $\pm$ 0.03	&0.44 $\pm$ 0.05	&0.54 $\pm$ 0.04	& \textbf{0.57} $\pm$ 0.02\\
\hline
$\Avg{h}$ & \quad	&0.17 $\pm$ 0.03	& 0.44 $\pm$ 0.06	& 0.22 $\pm$ 0.04\\
\hline
$CO$ & \quad	& \quad & 0.36 $\pm$ 0.01 & 0.44 $\pm$ 0.04 \\
\hline
$LRCO$ & \quad	& \quad	& \quad & 0.14 $\pm$ 0.05 \\
\end{tabular}
}
}
\caption{Classification performances of the classical measures and Matthews correlation coefficient with quadratic discriminant analysis.
The tables show the performances of couples of observables, with the performance of the single observables along the diagonal; the best performance is bold-typed.
The results presented are the average values of 10-fold cross-validation over 10000 instances and their standard deviation.}\label{OLDO}
\end{table*}

\section{Discussion}

Trying to deduce properties of the proteins from their structure is still an open challenge: in this paper we propose novel network observables based on the contact map for these purposes, in particular to discriminate between TS proteins that present only two configurations (folded/unfolded) and MS proteins with a richer landscape of stable and metastable states.

One of the new observables, $R_0$, simply counts  the density of inter-residue distances in which there are no contacts, but nonetheless it is very powerful for this classification purposes though being independent on protein size.
We remark that some known parameters used for TS-MS classification, such as the number of residues $N_C$, are ''extensive'' variables, thus, since many long proteins are MS and many small proteins are TS, it is very likely that these features have a limited discriminating power, in particular in the ''gray region'' of small MS and long TS proteins (as it is exactly the case in the analyses we have performed).
This means that the information contained in the PCN bands (the diagonals of the related adjacency matrix containing links between $d-$neighboring nodes) is very relevant, and possibly could be further explored by other measures.

Another class of observables we introduce is based on the eigenvalues of the Laplacian operator applied to PCNs.
In analogy with the physical Laplacian operator (acting on Euclidean space) the eigenvalues and eigenvectors can be put in relation with the main vibrational modes of the network and their respective frequencies.
We remark that for Laplacian observables it was important to emphasize the role of long-range residue contacts, that effectively characterize the protein 3D structure, by removing the protein backbone with a procedure that preserves the network connectivity as a unique component: based on PCN properties, this filtering of non-relevant links is protein-specific, differently from the more general definition of long-range interaction commonly used, with a unique threshold for all proteins. 
Our Discriminant Analysis showed that in general TS and MS proteins are better classified by larger Laplacian eigenvalues, corresponding to fast vibrating modes, at difference with small eigenvalues such as the Fiedler number. 
From the best performing couple of observables, $\lambda_{N-1}$ and $R_0$ see Fig. \ref{panel_class}, we deduce that TS proteins have larger values of fast-vibrating frequencies, and a larger number of inter-residue contacts (as can be seen in Fig. \ref{panel_class} A).
An interesting remark is that the vibrating modes associated to large eigenvalues tend to be more localized in specific residue chain regions (such as focusing modes in optics and whispering modes in acoustics \cite{ChenZhou93}).
It seems thus that the vibrating dynamics associated to specific regions of the residue may have a relevant role in these folding processes.

The other observable we introduced, based on the concept of network ensemble, depends on an estimation of the size of networks ensembles (from a canonical Statistical Mechanics point of view) that share common constraints (in our case the degree and the strength sequence of the PCN).
As expected by the physical meaning of entropy, that counts the number of ''microstates'' corresponding to a ''macrostate'' characterized by some fixed constraints, TS proteins show a smaller value of $S_R$, that can be interpreted as a smaller number of topological configurations available to the related networks. 
A high number of available PCN states, given a fixed degree and strength sequence, is thus very likely associated to proteins with more intermediate states during the folding process.


In conclusion, the high classification performance achieved, together with a direct physical interpretation, indicate that the newly introduced network-based observables can be relevant for a better comprehension of protein folding processes.





\end{document}